\documentclass [letterpaper,10pt]{article}

\twocolumn

\usepackage [dvips]{graphicx}

\usepackage[centertags]{amsmath}

\numberwithin{equation}{section}

\begin{document}

\begin{center}

\title{\normalsize{DYNAMICAL EFFECTS OF THE RADIAL GALACTIC TIDE ON AN OORT CLOUD OF COMETS FOR STARS WITH DIFFERENT MASSES
AND VARYING DISTANCES FROM THE GALACTIC CENTER} \\
\vspace{4mm}
Marco Masi \\
\vspace{2mm} \footnotesize{Dipartimento di Fisica G. Galilei, Padova, Italy.} \\
\vspace{1mm} \footnotesize{Email: marco.masi@spiro.fisica.unipd.it} \\
\vspace{4mm} \normalsize \quotation{\center{ABSTRACT} \\ \vspace{1mm} \textrm{We show that the
effects of radial galactic tides alone can change the dynamics of an Oort cloud of comets around a
star in a highly sensitive manner. The Oort cloud dynamics depends primarily from the star's
position in the Galaxy and its mass. The dynamical variation is characterized by a dramatic change
of comet perihelia, many of which are shifted towards the inner planetary system if the star's
position goes towards the galactic center or if its mass is set to be smaller. These preliminary
result suggested us a possible future line of research which aims to a better dynamical
generalization of comet orbits around extrasolar systems that might be tested through numerical
simulations of an Oort cloud of comets around non solar type stars and in different galactic or
extragalactic environments.}}}

\date{\vspace{-9mm} \small 9 October, 2002}

\maketitle

\end{center}

\hyphenation{go-ing}

\flushleft

\begin{center}
\small{1. INTRODUCTION}
\end{center}

\vspace{3mm}

\normalsize

\vspace{3mm}

The discovery of extrasolar planetary systems revealed a broad range of quite different features than those possessed
by our own solar system. This suggests what has long been suspected: our solar system is a special outcome of
planetary formation and evolution between the many existing elsewhere. Then, as it is well known, our sun has a mass
and spectral class which are only one of the many possible either. Moreover, the sun has also a specific position in
the Galaxy which, as it will become clear in this paper, isn't equivalent to any other.

\vspace{3mm}

Until recently, research on the formation and evolution of planetary systems focused on our own
solar system, that is, almost exclusively on a solar type star in the local galactic environment.
Formation of planets from an aggregation of planetesimals departing from a primordial nebula, and
the dynamical evolution of an Oort cloud of comets, has been investigated numerically for a
gravitational potential produced by a solar mass star under the galactic tidal perturbation
produced by our specific local galactic potential (i.e. that which results at about 8 Kpc from the
Galaxy center). In the last decade a line of research emerged with more general models of planetary
formation also for non solar type stars. However, still very few has been done to investigate the
dynamics of a cloud of comets for non solar mass stars moving in a different galactic environment
or stellar system by a comparative study with our own local Oort Cloud.

\vspace{3mm}

The aim of my thesis (Masi, 2001), was to study some aspects of how the local cometary dynamics depends on the
physical parameters and on the global dynamics of the Galaxy. This last one is strongly affected by the dark
matter halo which determines also the dimension of the bright halo and the central density of the Freeman disk and
the radial excursions of stellar orbits. Impelled by this relationship between global and local aspects we focused
our attention on the possible analogous connection between the global galactic gravitational potential and the
local Oort Cloud dynamics of cometary nuclei under a variation of its central star mass and orbit.

\vspace{3mm}

The effects of galactic tides on the Oort cloud dynamics has been studied extensively by many authors in nearly two
decades (e.g. Byl (1983), Bailey (1983), Heisler \& Tremaine (1986), Torbet (1986), Matese \& Whitman (1992) and
others). From some of these studies (e.g. Heisler \& Tremaine (1987), Matese \& Whitman (1995)) the hypotheses that
close stellar encounters and galactic tides produce comet showers towards the inner solar system becomes plausible.

\vspace{3mm}

However, all these studies considered exclusively our star, the sun, in the local galactic
environment. Moreover they made approximate analytical calculations or a statistical analysis
through Monte Carlo integrations and didn't investigate further the qualitative aspects of the
single exact comet orbit under the influence of the galactic force field. Whereas my approach was
characterized by another philosophy: due to a lack of computing power I integrated only very few
orbits, which for the purposes of this research line were considered particularly representative,
and renounced, at least in this initial study, to statistical quantitative estimates concentrating
my attention on the study of the orbital evolution of single test particles.

\vspace{3mm}

The emerging behaviour is quite complex and can't be exhausted briefly. The complexity of the
phenomenon is easily explained if we remember that we are studying a system which is described by
nonlinear differential equations and where the effects of chaotic dynamics can no longer be
neglected. Mentioning therefore here only the most relevant aspects, deferring a more detailed
analysis to future work, I nevertheless would like to outline roughly how radial tidal
perturbations (which is induced by the matter contained in the star's orbital radius) produce,
among other effects, an injection of comets towards the inner planetary system.

\vspace{3mm}

In $\S$ $2.1$ we first construct an approximate galactic potential. In $\S$ $2.2$ we show that a
variation of the orbital distance from the Galaxy center of a solar mass star influences comet
orbits in a significant way. In $\S \, 2.3$ we analyze to what extent the same dynamical evolution
depends from the star's mass, keeping instead the star's distance from the Galaxy center nearly
unaltered.

\vspace{3mm}

It is clear that this is only a very specific analysis. First of all we limited our numerical simulation on the
two dimensional galactic plane, and this means that we neglected the orthogonal disk tide which, in the outer
parts of the Galaxy, is stronger then the radial one\footnote{We choose to analyze radial galactic tides instead
of orthogonal ones, because we assume that these show a greater radial influence for a radial change.}. Moreover,
we did not consider the effect of close stellar encounters and of giant molecular clouds on comet orbits.
Furthermore, we used a set of only 48 particles which is clearly not enough for any statistical consideration. So,
finally what has been done here has nothing to do with a realistic representation of an Oort cloud for extrasolar
systems.

\vspace{3mm}

But, in this preliminary study, it was not our intention to develop a realistic Oort cloud model.
What we had in mind instead was to show, by an exact integration of few orbits, that the dynamical
evolution of comet orbits must be highly sensitive to the environment in which a star moves and to
the mass of the star itself. To highlight this it is not necessary to build a sophisticated
numerical model with many particles. Only an integration of a few dozens of orbits is amply
sufficient to show explicitly the existence and intensity of some dynamical effects. So, we
believe, what has been outlined has nevertheless an important qualitative significance. Including
other effects as stellar encounters and giant molecular clouds can only magnify the effects we are
describing here.

\vspace{3mm}

Anyhow, after this preliminary study, in $\S \, 3$, we will suggest a more elaborate algorithm and
a possible future line of research which aims to a more realistic model that might permit us to
make also more precise statistical predictions.

\vspace{3mm}

\setcounter{section}{2}

\begin{center}
\small{2. COMET ORBITS UNDER RADIAL GALACTIC PERTURBATION}
\end{center}

We investigated first the structure and distribution of matter in the Galaxy and the resulting galactic
gravitational potential. To do this beyond a certain amount of precision is not an easy task for two reasons.
First of all because the exact matter distributions, related to all the components of our Galaxy, are still not
very well known and secondly, if we would like to build a realistic potential, we should take in account many
irregularities and some asymmetries (for example the rotating triaxial non homogeneous matter distribution of the
bulge) which lead to mathematical and algorithmic complications. A better determination of matter distribution in
the Galaxy and a more realistic model of its gravitational potential needs further results from future research.

\vspace{3mm}

\setcounter{subsection}{1}

\begin{center}
2.1 \textsl{The construction of the galactic potential}
\end{center}

However, at this initial stage and as a first approximation, we built up a simple but representative galactic mass
distribution of the following components.

\vspace{3mm}

1) A central spherical bulge which, according to a Plummer model is represented analytically by a spherical
potential, $\Phi_{BG}(r)$, without diverging in the central regions:

\vspace{-0mm}

\small
\begin{displaymath}
\Phi_{BG}(r)= - \frac{G M_{BG}}{\sqrt{r^{2} + r_{c}^{2}}} \, ,
\end{displaymath}
\normalsize

where $r$ is the distance from the galactic center, $r_{c}$ the core radius and $M_{BG}$ the total mass of the
bulge (we choose $r_{c} = 420$ pc and $M_{BG} = 1.6 \times 10^{10} M_{\odot}$ (Flynn et al., 1996)).

\vspace{3mm}

2) The galactic disk has been modeled with a two dimensional Freeman disk (Freeman, 1970) by a mass column density
distribution as:

\vspace{-1mm}

\small
\begin{displaymath}
\Sigma(r) = \Sigma_{0} \, e^{-r/r_{d}} \, ,
\end{displaymath}
\normalsize

where $\Sigma_{0}$ is the central disk column density and $r_{d}$ is the disk's scale length (we choose, after a
comparative analysis of different existing models, like Schmidt's (1985), Van der Kruit's (1986) and other models:
$\Sigma_{0}= 492 \, M_{\odot} / pc^{2}$; $r_{d} = 3.5 \, Kpc$). This leads to a gravitational potential (Binney \&
Tremaine, 1987):

\vspace{-3mm}

\small
\begin{equation}
\hspace{-1mm}
\Phi_{d}(r) = - \pi G \Sigma_{0} \, r \, \left[ \, I_{0} \left( r/2 r_{d} \right) \, K_{1}(r/2 r_{d}) - I_{1}(r/2
r_{d}) \, K_{0}(r/2 r_{d}) \, \right] \, , \label{freeman}
\end{equation}
\normalsize

\vspace{0mm}

where $I_{n}$ and $K_{n}$ are modified Bessel functions.

\vspace{3mm}

3) A spherical pseudo-isotherm dark halo:

\vspace{-1mm}

\small
\begin{displaymath}
\rho_{D}(r) = \frac{\rho_{0}}{1 +  (\frac{r}{r_{H}}) ^{2}} \, ,
\end{displaymath}
\normalsize

with $\rho_{0}$ the central density and $r_{H}$ the scale length of the dark halo (we choose, because of
cosmological considerations made by Bullock et al. (2001), $\rho_{0}= 0.05 \, M_{\odot} / pc^{3}$, and in order to
recover a flat rotation curve, $r_{H}= 4.6 Kpc$). This is a spherical mass distribution which corresponding
gravitational potential can be recovered as (see Binney \& Tremaine , 1987, pag. 36):

\vspace{-0mm}

\small
\begin{displaymath}
\Phi_{D}(r) = -4\pi G\rho_{0}r_{H}^{2}\left[\frac{\ln(1 + r/r_{H})}{r/r_{H}} - \frac{1}{1 + R_{max}/r_{H}} \right]
\, ,
\end{displaymath}
\normalsize

\vspace{3mm}

where $R_{max}$ is the extension of the dark halo (this value however is not essential for our
considerations because, when calculating the component of the force field associated to
$\Phi_{D}(r)$, $\mathbf{F}_{D} = - \nabla \Phi_{D}$, $R_{max}$ falls out in the derivative).

\vspace{3mm}

If we pass from spherical to Cartesian galactocentric coordinates, through a substitution \small $r
= \sqrt{X^{2} +Y^{2} + Z^{2}}$ \normalsize, then the galactic gravitational potential at the
position in galactocentric coordinates $\left( X, Y, Z \right)$ of the test particle representing
the single comet nucleus, can be represented as:\footnote{$\Phi_{d}(X,Y,Z)$ is a three dimensional
gravitational potential for Freeman disks we will however not use here. It can be shown (Binney \&
Tremaine, 1987, p. 77) that in two dimensions, on the galactic disk, it reduces to \ref{freeman}.}

\vspace{-3mm}

\small
\begin{eqnarray}
\lefteqn{\Phi_{Gal}(X,Y,Z) = {} } \nonumber \\
& & {} \Phi_{BG}(X,Y,Z) + \Phi_{d}(X,Y,Z) + \Phi_{D}(X,Y,Z) \, . \nonumber
\end{eqnarray}
\normalsize

Adding the gravitational potential of a star, $\Phi_{\star}$, with galactocentric position
$(X_{\star},Y_{\star},Z_{\star})$ produced at $\left( X, Y, Z \right)$, which is given by:

\vspace{-3mm}

\small
\begin{eqnarray}
\lefteqn{\Phi_{\star}(X - X_{\star}, Y - Y_{\star}, Z - Z_{\star}) = {} } \nonumber \\
& & {} - \frac{G M_{\star}}{\sqrt{(X - X_{\star})^{2} + (Y - Y_{\star})^{2} + (Z - Z_{\star})^{2}}} \, ,
\nonumber
\end{eqnarray}
\normalsize

then the total potential $\Phi_{tot}$ in (X,Y,Z) becomes:

\vspace{-3mm}

\small
\begin{eqnarray}
\lefteqn{\Phi_{Tot}(X,Y,Z,X_{\star},Y_{\star},Z_{\star}) = {} } \nonumber \\
& & {} \Phi_{Gal}(X,Y,Z) + \Phi_{\star}(X - X_{\star}, Y - Y_{\star}, Z - Z_{\star}) \, . \nonumber
\end{eqnarray}
\normalsize

Since we are actually interested only in the galactic radial tides which act on a star when it is on the galactic
disk, we may neglect the small orthogonal excursions of the sun ($< 100$ pc, (Bash, 1986)) and can then reduce the
problem to a two dimensional one. By applying to the particle the usual equations of motion
$\mathbf{F_{\mathrm{pt}}} = - \nabla \Phi_{Tot}$, we obtain the differential systems:

\vspace{-3mm}

\small
\begin{eqnarray}
\ddot{X}(t) = - \frac{\partial \Phi_{Tot}}{\partial X} [X(t),Y(t),X_{\star}(t),Y_{\star}(t)]; \nonumber \\
\ddot{Y}(t) = - \frac{\partial \Phi_{Tot}}{\partial Y} [X(t),Y(t),X_{\star}(t),Y_{\star}(t)]; \nonumber
\end{eqnarray}
\normalsize

\vspace{-0mm}

\small
\begin{displaymath}
X(0) = R_{g} + R_{pt}; \qquad Y(0) = 0;
\end{displaymath}
\normalsize

\vspace{-0mm}

\small
$\hspace{23mm} \dot{X}(0) = 0; \hspace{18mm}  \dot{Y}(0) = V_{pt} \, , $
\normalsize

\vspace{3mm}

where (X(0), Y(0)) and $V_{pt}$ are respectively the initial position and speed of the particles, $R_{pt}$ the
initial distance of a particle from the star and $R_{g}$ the initial distance of the star from the galactic
center.

\vspace{3mm}

The star's orbit in time $(X_{\star}(t), Y_{\star}(t))$ has to be obtained firstly. This is done in a perfectly
analogous way through $\textbf{F}_{\star} = - \nabla \Phi_{Gal}$, that is:

\vspace{-3mm}

\small
\begin{eqnarray}
\ddot{X}_{\star}(t) = - \frac{\partial \Phi_{Gal}}{\partial X} [X(t),Y(t)]; \nonumber \\
\ddot{Y}_{\star}(t) = - \frac{\partial \Phi_{Gal}}{\partial Y} [X(t),Y(t)]; \nonumber
\end{eqnarray}
\normalsize

\vspace{-0mm}

\small
\begin{displaymath}
\hspace{-6mm}
X_{\star}(0) = R_{g} \, ; \qquad Y_{\star}(0) = 0 \, ;
\end{displaymath}
\normalsize

\vspace{-0mm}

\small
$\hspace{23mm} \dot{X}_{\star}(0) = 0; \hspace{10mm} \dot{Y}_{\star}(0) = V_{t_{\star}}\, ,$
\normalsize

\vspace{3mm}

with $V_{t_{\star}}$ the initial tangential velocity of the star (for our sun $V_{t_{\star}} \sim 220$ Km/s and
$R_{g} \sim 8$ Kpc).

\vspace{3mm}

After a numerical integration we obtain the trajectory of the particles in galactocentric coordinates which can
easily be transformed into the heliocentric ones:

\vspace{-0mm}

\begin{displaymath}
x(t) = X(t) - X_{\star}(t) \, ; \qquad y(t) = Y(t) - Y_{\star}(t) \, .
\end{displaymath}

\vspace{2mm}

\setcounter{subsection}{2}

\begin{center}
\textsl{2.2 The Oort cloud dynamics dependence from its star's orbit}
\end{center}

\vspace{0mm}

The first part of our study visualizes the dynamic effects of radial galactic tides on a set of particles for a star
with a fixed mass but each time with a different orbit around the Galaxy center for each simulation.

\vspace{3mm}

One of our aims was to show that the effects of galactic radial tides alone, which aren't
particularly important at the sun's distance from the galactic center ($\sim 8$ Kpc), become
relevant as soon as the distance decreases and get strong enough to inject nearly parabolic comets
towards the inner planetary system. The result is achieved by a numerical interpolation of highly
elliptic orbits of test particles around a star with solar mass moving on a nearly circular
galactic orbit\footnote{We remind that in a galactic potential closed circular or elliptic orbits
do not exist, also in the case we would neglect the stochastic interactions with other stars and
interstellar clouds. It would be more precise to regard stellar orbits on the galactic disk as
those resulting from a spherical gravitational potential: rosetta-like orbits.}, but with different
mean distances from the galactic center for each integration.

\vspace{3mm}

The comets which are most sensitive to the perturbative forces which can become visible in
planetary regions are those with greatest aphelia and relatively little perihelia. Moreover, the
efficiency of a tidal force depends also from the orientation of the comet's orbit semi-major axes
with respect to the galactic center. Therefore we have studied a particular set of 48 particles,
all with initial perihelion $q = 2000$ AU, different aphelia ($a = 40.000$ AU, $60.000$ AU,
$80.000$ AU, $100.000$ AU, $120.000$ AU, $140.000$ AU) and different galactic longitudes of their
semi-major orbital axis (l = $0^{\circ}$, $45^{\circ}$, $90^{\circ}$, $135^{\circ}$, $180^{\circ}$,
$225^{\circ}$, $265^{\circ}$, $315^{\circ}$). These orbital parameters represent a quite extreme
sample of the sun's Oort's cloud comets, and they are considered to represent some of the most
external comets still bound to the gravitational field of the sun and perturbed by external
galactic forces. Nevertheless they still represent a plausible set of external test particles which
emerge after the sun's Oort evolutionary history (Fernandez (1980), Weissman (1985), Remy \&
Mignard (1985)). If this still holds also for comet clouds around other stars in our Galaxy which
experienced different physical conditions and had almost certainly a different dynamic and
evolutionary history, remains an open question. As long as we will not execute more sophisticated
and realistic simulation models (and the aim of this paper is just to encourage such line of
research!), we will however continue to use this sample because it can give us at least a
qualitative idea about the order of magnitude of the sensitivity of an outer shell of the Oort
cloud under a variation of its central star mass and its position in the Galaxy.

\vspace{3mm}

The result of our numerical simulation can be observed in Fig. 1 through Fig. 4 \footnote{Some
graphs show segmented orbits. This is not because of coarse numerical calculation but only due a
time-stepped graphical output which becomes coarse for the increasing velocity at the perihelia.
The artifacts in the images are due compression.}

\vspace{3mm}

Fig. 1a: The galactic potential has been turned off and we obtain the usual Keplerian elliptic
orbits, as expected.

\vspace{3mm}

Fig. 1b: A closer view at the central internal regions up to 3000 AU.

\vspace{3mm}

Fig. 2a: The galactic potential (for a distance from the galactic center of 8 Kpc) has now been
turned on. Keplerian orbits are apparently only slightly perturbed by the galactic radial tide.

\vspace{3mm}

Fig. 2b and 2c: However the two closeup views at 3000 AU (compare this again with Fig. 1b) and 100
AU show another truth. The perihelia with an initial value of 2000 AU are injected towards the
inner regions, down to 85 AU, because of the sole galactic radial tidal perturbations.

\vspace{3mm}

Fig. 3a: The distance from the galactic center is 6 Kpc. The orbits get now still more "blurred".

\vspace{3mm}

Fig. 3b: Many comet perihelia literally invade the central regions of the Oort cloud.

\vspace{3mm}

Fig. 3c: Zooming in the inner planetary system we see the outer comets of the sample reach $1.5$ AU
(i.e. nearly the Earth orbit radius).

\vspace{3mm}

Fig. 4a: The distance from the galactic center is 2 Kpc, we are now in the bulge. The outer part of the cloud (those
particles with $100.000$ AU $< a < 140.000$ AU) becomes disrupted because of tidal stripping effects: the Roche's
tidal limit radius reaches the Oort cloud.

\vspace{3mm}

Fig. 4b and 4c: The inner regions become even more crowded by the remaining test particles.\footnote{ The
innermost regions ($0$ AU $< a < 100$ AU) seem to be again devoid of comet perihelia. However note that this is
only because in our simulation no particles with $80.000$ AU $< a < 100.000$ AU where present.}

\vspace{3mm}

In general note how for comets orbits (especially for the two most external aphelia of 120.000 and 140.000 AU)
with a perpendicular semi-major axis relative to the sun-Galaxy center line ($90^{\circ}$ galactic longitude),
radial tidal forces tend to squeeze the semi-minor axes injecting the perihelia towards the central planetary
system whereas those orbits parallel to that line ($0^{\circ}$ galactic longitude), experience the opposite
effect, that of shifting the perihelia towards the external regions, stripping them off more easily.

\setcounter{subsection}{3}

\begin{center}
\textsl{2.3 The Oort cloud dynamics dependence from the star's mass}
\end{center}

\vspace{0mm}

The second part of the study visualizes the dynamic effects of radial galactic tides on a set of particles for a star
on a fixed nearly circular orbit around the Galaxy center, but with a different mass for each integration. This means
that in this subsection we execute the same exact integration we have done in the previous one with the only
difference that this time we vary the star's mass. In our case we choose the sun's orbit about the galactic center at
about 8 Kpc. The resulting graphs are analogous to Fig. 1 $\div$ 4 but the perihelia injection is caused by a
variation of the star's mass instead of its position in the Galaxy.

\vspace{3mm}

Fig. 5a, 5b and 5c: The star's mass is $0.8 M_{\odot}$. This situation might be roughly compared with that of Fig.3.

\vspace{3mm}

Fig. 6a, 6b and 6c: The star's mass is $0.2 M_{\odot}$. Again, first effects of tidal stripping become apparent.

\vspace{3mm}

Fig. 7a and 7b: The star's mass is $2 M_{\odot}$. As expected, the stronger gravitational field
shields the galactic gravitational field better than in the solar mass case of Fig. 2. Orbits are
nearly Keplerian and the injection of perihelia is less efficient.

\vspace{3mm}

\setcounter{section}{3}

\begin{center}
3. CONCLUSION AND A PROPOSAL FOR A POSSIBLE FUTURE LINE OF RESEARCH
\end{center}

\vspace{0mm}

We consider this preliminary study sufficient to conclude that an environment dependent comet dynamics as such
exists and is characterized especially by a partial injection of comet perihelia towards the inner planetary system.
Simulating only the effects of radial galactic tides alone reveals a great change in cometary orbits. Including
other effects (like stellar encounters, orthogonal tides, giant molecular clouds, etc.) can only enforce this
conclusion.

\vspace{3mm}

Therefore I believe that it would be interesting to build a more realistic simulation model which
could tell us something more about the aspects of the evolution and dynamics of an Oort-like cloud
of comets in more general astrophysical contexts than those previously tested, i.e. for non solar
type stars in other gravity force fields than our local galactic environment, taking into account
other dynamical effects we neglected in this preliminary test. This could provide us some
information about the dynamical evolution and history of comet clouds in extrasolar systems.

\vspace{3mm}

For instance we might expect that in the inner parts of the Galaxy, stars have a much more chaotic and smaller
cloud of comets than those stars with the same mass at galactic periphery, because of the more intense
perturbations existing in the central galactic regions. Then, comet clouds around stars with lower (greater) mass
are more (less) sensitive to external perturbations than solar mass ones and it might be interesting to see how
and to what extent this mass parameter influences the history and final structure and dynamics of an Oort cloud.
Another aspect in our opinion worth to be investigated may also be the question of how a cloud of comets differs
for two identical stars (say as our sun) but one moving in a spiral galaxy and the other orbiting in a globular
cluster or other stellar systems?

\vspace{3mm}

Some evidences (e.g. Melnick et al. 2001)\footnote{However many stars are binary. Are comet clouds impossible in
binary systems or does some particular equilibrium configuration or resonance exist?} suggest that also other
stars have an Oort cloud . But we must also consider the possibility that around many stars the Oort cloud may not
form at all. For example in the inner regions of the Galaxy, stellar encounters are possibly so close and frequent
that they may be, on the long run, strong enough to destroy, due to gravitational stripping, the Oort cloud
entirely. Is the formation of an Oort cloud a rule or an exception?

\vspace{3mm}

Moreover, there is perhaps also an exobiological aspect in all that. It is now an established fact
that our terrestrial environment experienced impacts with asteroids and comets in its past
evolutionary history which possibly led to mass extinctions and profound geo-chemical changes on
our planet's surface. A too intense comet shower could have had disastrous consequences for the
emergence of life. As we have already shown, radial galactic tidal forces alone are sufficient to
inject comets of the outer part of the Oort cloud towards the inner planetary system. Fortunately,
at the sun's distance from the galactic center ($\sim 8$ Kpc) these effects are too weak to
determine a comet flux that might lead to too frequent impacts with our planet. But already at the
distance of 6 Kpc these effects can no longer be neglected and perhaps, in these regions a
biological evolution might have been affected somehow, whereas in the most inner parts of the
bulge, one might be tempted to say that life can even be threatened by continuous comet impacts. On
the other side there is some possibility that comets might have played an important role in
furnishing the primordial biological material necessary for the emergence of life on earth. A
complete absence of comet impacts (probably that's the case for planetary systems around stars in
peripheric galactic regions where star encounters, and giant molecular clouds are rare and tidal
forces weaker) might have resulted in a lack of a replenishment of some necessary biochemical
substances with inevitable consequences in terms of biological evolution. So, the existence of life
in a planetary system might possibly be somehow intrinsically related to the dynamical evolution of
the Oort cloud, which in turn depends from mass, position and orbit of a star in a Galaxy. If so,
this would imply the existence of "Galactic Habitable Zones" on the same line of Gonzalez et al.
(2001).

\vspace{3mm}

On the other side there is an interesting aspect about what rule cometary impacts might have played
in developing oceans on earth. Chyba (1987) proposed comets as the primordial source for water on
earth. But Blake et al. (1999) showed that, if we compare isotopic abundances of terrestrial with
cometary water, then it appears unlikely that more than 20\% of our ocean's water had cometary
origins. This suggests that for our peripheric position in the Galaxy, comets might have played
only a minor role in forming oceans. With the exception of Jupiter's moon Io (an perhaps Callisto
and Ganymede) only earth possesses oceans. But things might change dramatically for planetary
systems in the Galaxy's bulge. Here the probably huge intensity of comet showers on the surface of
some or all their planets might have been so strong to form giant oceans, or "Water Worlds", on
their surface which subsequently acquired the double function as a protective shield from other
outer impacts and as a candidate place for biological evolution.

\vspace{3mm}

Of course, at this stage, these are only mere hypothesis. But, in order to go beyond these
speculations, it is just for this reason that we suggest to extend the research on the Oort cloud
dynamics for extrasolar systems through a much more elaborated and realistic model, capable to shed
some light at least on some of the dynamical aspects related with these and many other questions.

\vspace{3mm}

I suggest to proceed as follows.

\vspace{3mm}

1) A more realistic three dimensional model of the galactic mass distribution, and therefore of the gravitational
force field of our Galaxy, from up to date astrophysical data has to be investigated. What do we know about the
mass of the central black whole, the mass and dimensions of the galactic nucleus, the triaxial mass distribution of
the bulge, the parameters of the Freeman-like galactic disk, the spheroid, the dark halo, the visible halo, the
distribution of giant molecular clouds, etc.? The correct determination of these components is essential for the
construction of a realistic galactic mass distribution and gravitational potential.

\vspace{3mm}

2) The statistical initial distribution of comets in phase space, with their orbital parameter
(perihelion, aphelion, inclination, latitude and longitude of their semi-major axes) has to be
determined. The present state of a cloud of comets depends from its past evolution which in turn
depends, among other things, again from the star's mass and its galactic position. The sun's
present Oort cloud is not representative as an initial condition for a simulation. Therefore, what
we have to do is to simulate the dynamic evolution of these comets from the time of the star's
protoplanetary creation, that is from their formation in the planetary region (the trans-Uranian
region in the case of the solar system), taking eventually in account a planetary migration
phenomenon\footnote{A mechanism of migration of the outer planets seems likely, especially after
the discovery of very massive planets in very close orbits around other stars. As B. M. S. Hansen
(2000) points out, it seems that planetary migration can be considered as another cause of failure
in developing an Oort cloud.} or collisional processes (Stern \& Weissman (2001)). So a preliminary
theoretical study of the most realistic comet distribution at formation time of the protoplanetary
cloud is needed.

\vspace{3mm}

3) A development of an optimized algorithm and its software implementation which includes planetary
perturbations, impacts of comets with the sun or other planets, perturbations with giant molecular
clouds, gravitational effects of close stellar encounters, elimination of particles because of
tidal stripping and other effects, has to be done.

\vspace{3mm}

4) Execution of the algorithm. Different runs are necessary in order to gain a realistic view. How does the dynamics
of the Oort cloud differ with varying distances from our galactic center? What if a star's orbit is highly eccentric
or is inclined to the galactic plane? How do comet orbits evolve around a sun mass star in other stellar systems like
a globular cluster? How does all this vary with varying stellar masses or planets position and mass? Else?

\vspace{3mm}

5) A study is necessary of how to evaluate correctly the final resultant output from the simulation and in the most
compact form (a huge amount of data is expected for each run). A statistical global analysis and representation of
the orbital parameters in time of the Oort cloud as a whole must be developed.

\vspace{3mm}

All this is perhaps a too ambitious study for a single researcher. However, a research group could do so, by
developing a software package, i.e. a sort of "Oort cloud simulator", ideally running it on a dedicated Beowulf
cluster system, capable to serve as a universal numeric simulation testing system for the study of the Oort cloud
dynamics in different astrophysical conditions. This would give us the possibility to study complex extrasolar
dynamical phenomena otherwise not possible to investigate.


\vspace{3mm}

\begin{thebibliography}{99}

\bibitem{Bailey} Bailey J. The effect of the galaxy on cometary orbits. Earth Moon Planets, 36, 263-273 (1983).

\bibitem{Bash} F. Bash in: The Galaxy and the solar System. The University of Arizona Press, pag. 35 Tucson (1986).

\bibitem{Tremaine} Binney, James. Tremaine, Scott. Galactic Dynamics. Princeton Series in Astrophysics (1987).

\bibitem{Blake} G. Blake, C. Qi, M.R. Hogerheijde, M.A. Gurwell, D.O.  Muhleman. Sublimation from icy jets as a probe
of interstellar volatile content of comets. Nature 398,213 (1999).

\bibitem{Bullock} J.S. Bullock, T.S. Kolatt, Y.Sigard, R.S.Sommerville, A.V.Kravtsov, A.A. Klypin, J.R.Primack e A.Dekel. Profiles of dark haloes: evolution, scatter
and environment. MNRAS 321,559-575 (2001).

\bibitem{Byl} Byl, J. Galactic perturbations of nearly-parabolic orbits. Moon and Planets 29, 121-137 (1983)

\bibitem{Chyba} C. F. Chyba The cometary contribution to the oceans of primitive Earth. Nature 330,632 -635 (1987).

\bibitem{Fernandez} Fernandez, Julio A. Evolution of Comet Orbits under the Perturbing Influence of the Giant Planets and
Nearby Stars. ICARUS 42, 406-421 (1980).

\bibitem{Gonzalez} Gonzalez G., Brownlee D., Ward P. The Galactic Habitable Zone I. Galactic Chemical Evolution.
astro-ph/0103165. Or: Refuges for Life in a Hostile Universe, Scientific American issue, October, 2001.

\bibitem{Freeman} Freeman, K.C. (1970), Astroph. J. 160,881.

\bibitem{Flynn} Flynn C., Sommer-Larsen J. e Christensen P.R. MNRAS 281,1027, (1996).

\bibitem{Hansen} Hansen, Brad M. S. Failed Oort Clouds and Planetary Migration. astro-ph/0004058

\bibitem{Heisler2} Heisler Julia, Tremaine Scott. The influence of the galactic tidal field on the Oort comet
cloud. Icarus, 13-26 (1986)

\bibitem{Heisler} Heisler Julia, Tremaine Scott, Alcock Charles. The frequency and intensity of comet showers
from the Oort cloud. Icarus 70, 269-288 (1987)

\bibitem{Masi} Masi Marco, Dissertation thesis on "The decentralization of the solar system and dynamical aspects of
the Galaxy", 2001, University of Padua, faculty of physics, Italy.

\bibitem{Whitman} Matese J. John, Whitman G. Patrick, A model of the galactic tidal interaction with the Oort comet
cloud. Celestial mechanichs and dynamical astronomy, 54:13-35 (1992).

\bibitem{Mateseetal} Matese J. John, Whitman Patrick G., Innanen Kimmo A., Valtonen Mauri J. Periodic modulation
of the Oort cloud comet flux by the adiabatic changing galactic tide. Icarus 116, 255-268 (1995).

\bibitem{Melnick} G.J. Melnick, D.A. Neufeld, K.E.S. Ford, D.J.  Hollenbach, M.L.N. Ashby. Discovery of water vapour
around IRC+10216 as evidence for comets orbiting another star. Nature 412,160-163 (2001).

\bibitem{Remy} Remy,F., Mignard,F. Dynamical evolution of the Oort cloud. I. A Monte carlo simulation.
Icarus 63, 1-19, (1985).

\bibitem{Schmidt} Schmidt, Maarten. Models of the Mass Distribution of the Galaxy in The Milky Way Galaxy, H. van
Woerden et al. (eds.), IAU Symposium (1985).

\bibitem{Weissman2} Stern S., Weissman Paul R. Rapid Collisional evolution of comets during the formation of the
Oort cloud. Nature, Vol. 409, Issue 6820, pp.589-591 (2001).

\bibitem{Torbett} Torbett, Michael V. Dynamical Influence of Galactic Tides and Molecular Clouds on the Oort Cloud of Comets.
In: The Galaxy and the solar System. The University of Arizona Press, Tucson (1986).

\bibitem{Kruit} van der Kruit, Pieter C. Surface Photometry of edge on Spiral Galaxies. A. \& A. 157, 230-244 (1986).

\bibitem{Weissman} Weissman Paul R. Dynamical evolution of the Oort Cloud. In Dynamics of Comets, ed. A. Carusi e
G.B. Valsecchi (Dordrecht: D. Reidl), pag.87-96, (1985).

\end{thebibliography}
\end{document}